# Superconductivity in LaPd$_2$Bi$_2$ with CaBe$_2$Ge$_2$-type structure

Qing-Ge Mu[1,2], Bo-Jin Pan[1,2], Bin-Bin Ruan[1,2], Tong Liu[1,2], Kang Zhao[1,2], Lei Shan[1,2,3], Gen-Fu Chen[1,2,3], and Zhi-An Ren[1,2,3,*]

[1] Institute of Physics and Beijing National Laboratory for Condensed Matter Physics, Chinese Academy of Sciences, Beijing 100190, China

[2] School of Physical Sciences, University of Chinese Academy of Sciences, Beijing 100049, China

[3] Collaborative Innovation Center of Quantum Matter, Beijing 100190, China

* Email: renzhian@iphy.ac.cn

**Abstract**

Here we report the synthesis and superconductivity of a novel ternary compound LaPd$_2$Bi$_2$. Shiny plate-like single crystals of LaPd$_2$Bi$_2$ were first synthesized by high-temperature solution method with PdBi flux. X-ray diffraction analysis indicates that LaPd$_2$Bi$_2$ belongs to the primitive tetragonal CaBe$_2$Ge$_2$-type structure with the space group $P4/nmm$ (No. 129), and the refined lattice parameters are $a$ = 4.717(2) Å, $c$ = 9.957(3) Å. Electrical resistivity and magnetic susceptibility measurements reveal that LaPd$_2$Bi$_2$ undergoes a superconducting transition at 2.83 K and exhibits the characteristics of type-II superconductivity. The discovery of superconductivity in LaPd$_2$Bi$_2$ with CaBe$_2$Ge$_2$-type structure may help to further understand the possible relationship between the occurrence of superconductivity and the crystal structures in 122-type materials.

## 1. Introduction

Since the discovery of superconductivity in LaFeAsO$_{1-x}$F$_x$, the high-$T_c$ iron-based superconductors have been extensively studied from both experimental and theoretical viewpoints [1-8]. However, the mechanism of the unconventional superconductivity is still to be resolved. To address such issues, numerous 3$d$, 4$d$, 5$d$ transition metal pnictide, silicide, germanide, chalcogenide materials crystallizing in the similar crystal structure with iron pnictide/selenide superconductors were studied [9-24]. Among of them, the 122-type compounds have attracted more attention due to the ease of synthesizing high-quality single crystals.

The 122-type materials mainly crystallize in the two different ThCr$_2$Si$_2$-type and CaBe$_2$Ge$_2$-type crystal structures, and both of them are variants of the BaAl$_4$-type structure [25]. The 122-family of iron-based superconductors belongs to the body-centered tetragonal ThCr$_2$Si$_2$-type structure [26-29]. The stacking sequence along $c$-axis in ThCr$_2$Si$_2$-type structure is Th-Cr$_2$Si$_2$-Th-Cr$_2$Si$_2$-Th, in which the chromium atoms are arranged in a square planar lattice with silicon atoms on either side of the chromium layer forming edge-sharing CrSi$_4$ tetragon. However, for the CaBe$_2$Ge$_2$-type structure, Be$_2$Ge$_2$ layer consisting of edge-sharing BeGe$_4$ tetragon and Ge$_2$Be$_2$ layer consisting of edge-sharing GeBe$_4$ tetragon stack along $c$-axis alternately. Till now, tens of the 122-type superconductors have been discovered within different transition metal based compounds [18-23,30-35]. Most of them are in ThCr$_2$Si$_2$-type structure, while only a few superconductors belong to CaBe$_2$Ge$_2$-type structure.

Pd-based 122-type pnictide compounds have been intensively studied by many groups, and some of them exhibit superconductivity at low temperatures [14,34-41]. Most of the Pd-based superconductors crystalize in ThCr$_2$Si$_2$-type structure, such as LaPd$_2$P$_2$ ($T_c$ ~ 0.96 K) [38], LaPd$_2$As$_2$ ($T_c$ ~ 1 K) [37], CaPd$_2$As$_2$ ($T_c$ ~ 1.27 K) [36], SrPd$_2$As$_2$ ($T_c$ ~ 0.92 K) [36], BaPd$_2$As$_2$ ($T_c$ ~ 3.85 K) [39], and high temperature phase SrPd$_2$Sb$_2$ ($T_c$ ~ 0.6 K) [41], and a few CaBe$_2$Ge$_2$-type structure Pd-based superconductors have been reported, such as low temperature phase SrPd$_2$Sb$_2$ ($T_c$ ~ 1.95 K or 1.4 K), LaPd$_2$Sb$_2$ ($T_c$ ~ 1.3 K) and SrPd$_2$Bi$_2$ ($T_c$ ~ 2 K) [34,35,40,41]. Therefore, Pd-based 122-type materials provide a rare platform for the study on the possible relationship between the occurrence of superconductivity and the crystal structures in 122-type materials.

Here we report a new $CaBe_2Ge_2$-type compound $LaPd_2Bi_2$. Electrical resistivity and magnetic susceptibility measurements indicate that it exhibits superconductivity with $T_c$ 2.83 K.

## 2. Experimental details

The single crystals of $LaPd_2Bi_2$ were grown by the high-temperature solution method using PdBi as the flux. First, the PdBi precursor was synthesized with the conventional solid-state-reaction method using high purity elemental Pd powder (99.99%) and Bi powder (99.999%) as the starting materials. The stoichiometric mixture of Pd and Bi was pressed into pellets and loaded into an alumina crucible, followed by being sealed into an evacuated quartz tube. The sealed quartz tube was sintered at 1073 K for 30 h. Second, $LaPd_2Bi_2$ single crystals were grown with La pieces (99.5%) and PdBi flux. The La pieces and PdBi powder were mixed in a molar ratio of 1:4, loaded into an alumina crucible, and then sealed into an evacuated quartz tube. The sealed quartz tube was heated to 1273 K and sintered for 24 h, followed by fast cooling down to 1073 K in 1 h and then slowly cooling down to 923 K at a rate of 1 K/h. Last, the shiny plate-like single crystals with a typical dimension 0.5×0.5×0.03 $mm^3$ were separated by decanting the flux with a centrifuge at 953 K. To avoid contamination, all the weighing, mixing, pressing and grinding processes were performed in a glove box ($O_2$, $H_2O$ < 1 ppm) filled with high purity Ar.

The crystal structure was characterized by powder x-ray diffraction (XRD) at room temperature from 10° to 80° with a PAN-analytical x-ray diffractometer using Cu-$K_\alpha$ radiation. The chemical composition of the single crystals was analyzed with both Phenom scanning electron microscope (SEM) with an energy-dispersive spectrometer (EDS) and inductively coupled plasma-atomic emission spectroscopy (ICP). The electrical resistivity was measured with a Quantum Design physical property measurement system (PPMS) by the standard four-probe method. The temperature (or field) dependence of dc magnetization was measured with a Quantum Design magnetic property measurement system (MPMS).

## 3. Results and discussion

The obtained $LaPd_2Bi_2$ single crystals are plate-like and fragile. Figure 1(b) shows the typical SEM morphology of $LaPd_2Bi_2$ single crystals. The XRD patterns for the crushed single crystals and the surface of a piece of single crystal were collected and shown in Fig. 1. The XRD patterns are well indexed with the primitive tetragonal $CaBe_2Ge_2$-type structure without any peaks arising from impurity, which indicates that the single crystals of $LaPd_2Bi_2$ are single phase. The powder XRD data for the crushed single crystals were analyzed with GSAS using Rietveld refinement, and the refined lattice parameters are $a$ = 4.717(2) Å, $c$ = 9.957(3) Å with the space group $P4/nmm$ (No. 129). The lattice structure expands slightly along $a$-axis, while shrinks a little along $c$-axis contrasting with $LaPd_2Sb_2$ which has the same crystal structure [34]. The XRD patterns on the surface of the single crystals only show the peaks of (00l), suggesting that the single crystals are well oriented. The refined atomic coordinates are listed in table 1, and the crystal structure of $LaPd_2Bi_2$ is displayed in Fig. 1(a). ICP measurements reveal that the chemical composition of the as-grown $LaPd_2Bi_2$ single crystals is $LaPd_{1.84}Bi_2$, indicating that Pd vacancies exist in the crystal structure, which is also confirmed by the results of EDS measurements on the surface of single crystals and XRD analysis.

The temperature dependence of in-plane electrical resistivity was measured from 1.8 to 300 K at zero field as shown in Fig. 2(a), which exhibits metallic behavior in the normal state and undergoes a superconducting transition at low temperature. The onset $T_c$ is 2.83 K which is clearly displayed in the inset, and it is higher than the values of $LaPd_2P_2$, $LaPd_2As_2$ and $LaPd_2Sb_2$ [34,37,38]. The superconducting transition is sharp with the width $\Delta T$ about 0.08 K. The residual resistivity ratio (RRR) is only about 1.2, which can be explained by the increased electron scattering arising from the Pd vacancies. Similar phenomenon was also observed in $LaPd_{1-x}Bi_2$ [42]. To further study the superconductivity in $LaPd_2Bi_2$, the upper critical field was measured and shown in Fig. 2(b). The inset shows the temperature dependence of in-plane electrical resistivity from 1.8 to 3.1 K at various magnetic fields with the field perpendicular to the $ab$-plane. Obviously, $T_c$ shifts to lower temperature gradually with the magnetic field increasing from 0 to 3.5 kOe, and superconductivity is suppressed when the field is above 3.5 kOe. The temperature dependence of upper critical field deduced from the data in the inset is displayed in Fig. 2(b), in which $T_c$ under different magnetic fields is determined as illustrated in the inset of Fig. 2(a). The data are well fitted with the Ginzburg-Landau (GL) formula, $H_{c2}(T) = H_{c2}(0)(1 - t^2)/(1 + t^2)$, where $t = T/T_c$, and the calculated upper critical field is 9.4 kOe.

To confirm the bulk superconductivity, the temperature dependence of dc magnetic susceptibility was measured from 1.8 to 3.5 K with the magnetic field perpendicular to the *ab*-plane. The data from both zero-field-cooling (ZFC) and field-cooling modes (FC) under a magnetic field of 2 Oe are shown in Fig. 3(a). Both of the data display significant diamagnetic transition below 2.7 K, which is consistent with the electrical resistivity measurements. The superconducting volume fraction from ZFC data at 1.8 K is close to 100%, suggesting the bulk superconductivity of $LaPd_2Bi_2$. Figure 3(b) shows the isothermal magnetization at 1.8 K, which exhibits the characteristics of type-II superconductors.

In summary, we successfully synthesized the single crystals of $LaPd_2Bi_2$ which crystalize in the $CaBe_2Ge_2$-type crystal structure by the high-temperature solution method using PdBi as the flux. Electrical resistivity and dc magnetic susceptibility measurements confirm that $LaPd_2Bi_2$ shows type-II superconductivity at 2.83 K. The discovery of superconductivity in $CaBe_2Ge_2$-type $LaPd_2Bi_2$ provides a new platform for the study on the possible underlying relationship between the superconductivity and crystal structures in 122-type compounds.

**Acknowledgements**

The authors are grateful for the financial supports from the National Natural Science Foundation of China (No. 11474339 and 11774402), the National Basic Research Program of China (973 Program, No. 2016YFA0300301) and the Youth Innovation Promotion Association of the Chinese Academy of Sciences.

**Figure and table captions:**

**Figure 1:** (Color online) (a) The $CaBe_2Ge_2$-type crystal structure of $LaPd_2Bi_2$. (b) The typical SEM morphology of single crystals. (c) The XRD pattern on the surface of a piece of single crystal. (d) The powder XRD pattern of the crushed single crystals, which was refined with Rietveld analysis using GSAS.

**Figure 2:** (Color online) (a) The temperature dependence of in-plane electrical resistivity, and the inset is the enlarged view around $T_c$. (b) The temperature dependence of upper critical field, which is well fitted with GL formula. The inset shows the temperature dependence of in-plane electrical resistivity under various magnetic fields from 0 to 3.5 kOe around $T_c$.

**Figure 3:** (Color online) (a) The temperature dependence of dc magnetic susceptibility with ZFC and FC modes under a magnetic field of 2 Oe. (b) The isothermal magnetization at 1.8 K.

**Table 1:** Atomic coordinates of the crystal structure of $LaPd_2Bi_2$ from the refinement of powder XRD patterns for the crushed single crystals.

**Figure 1**

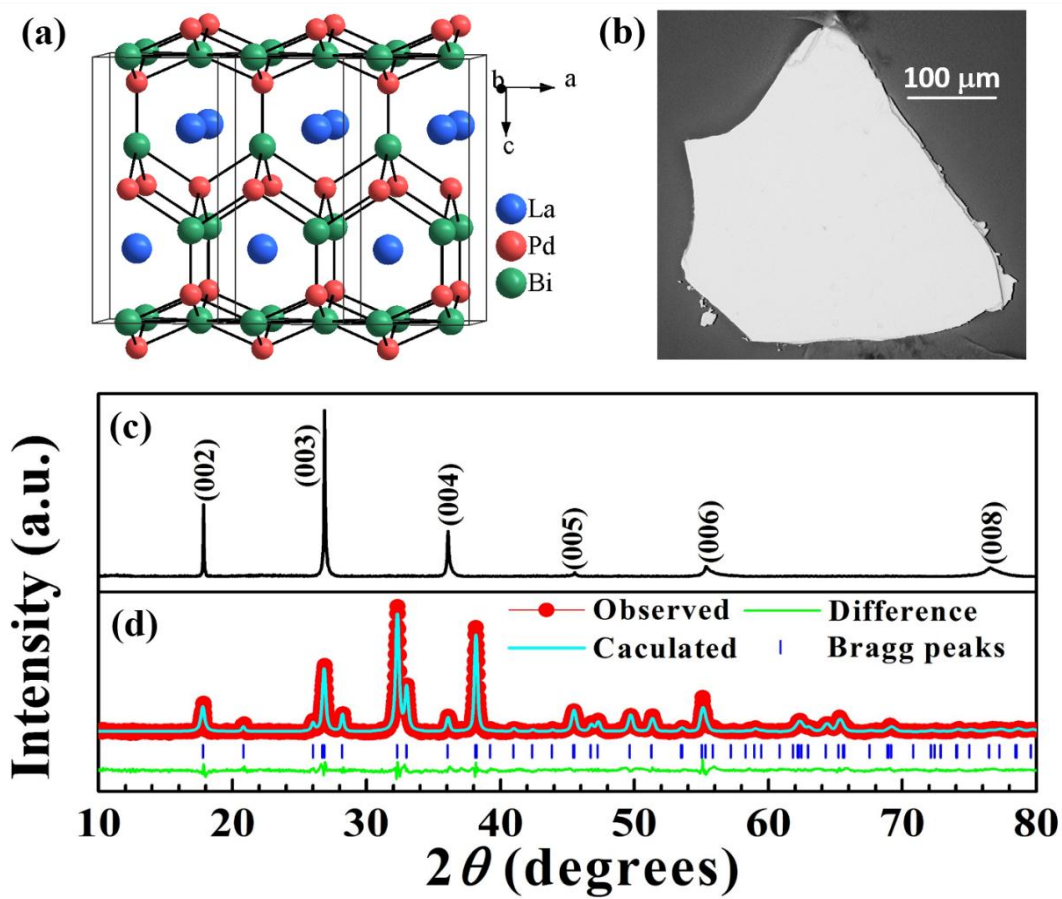

**Figure 2**

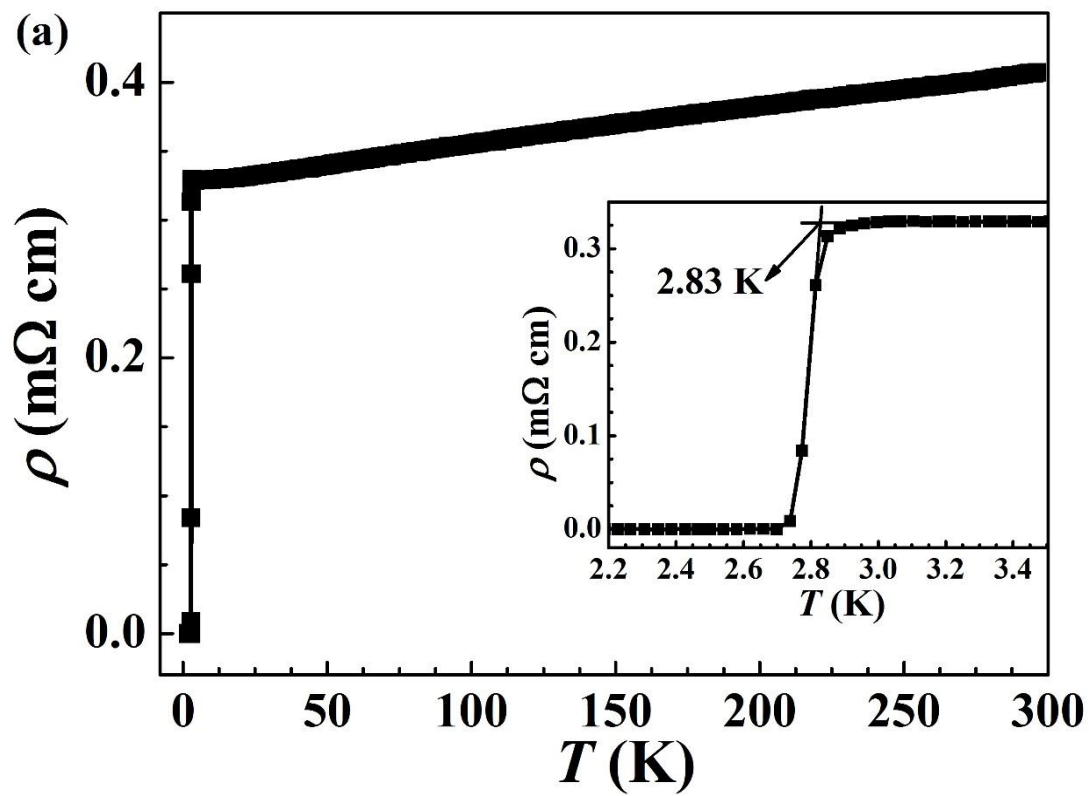

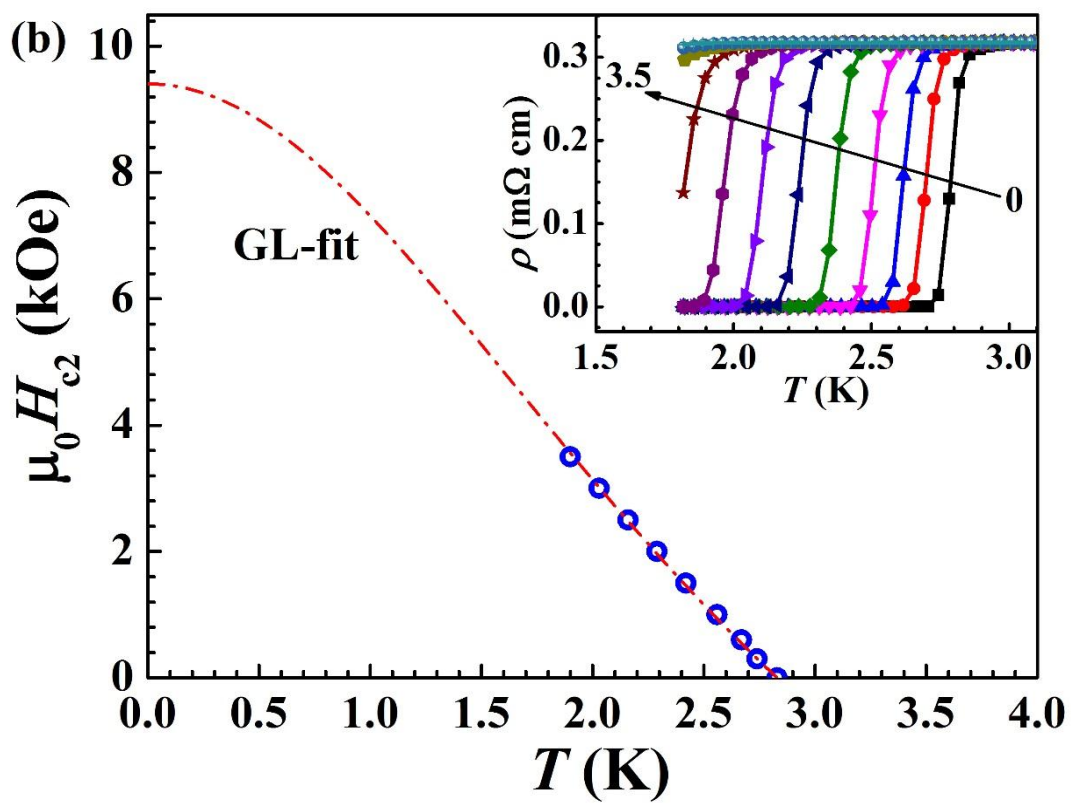

Figure 3

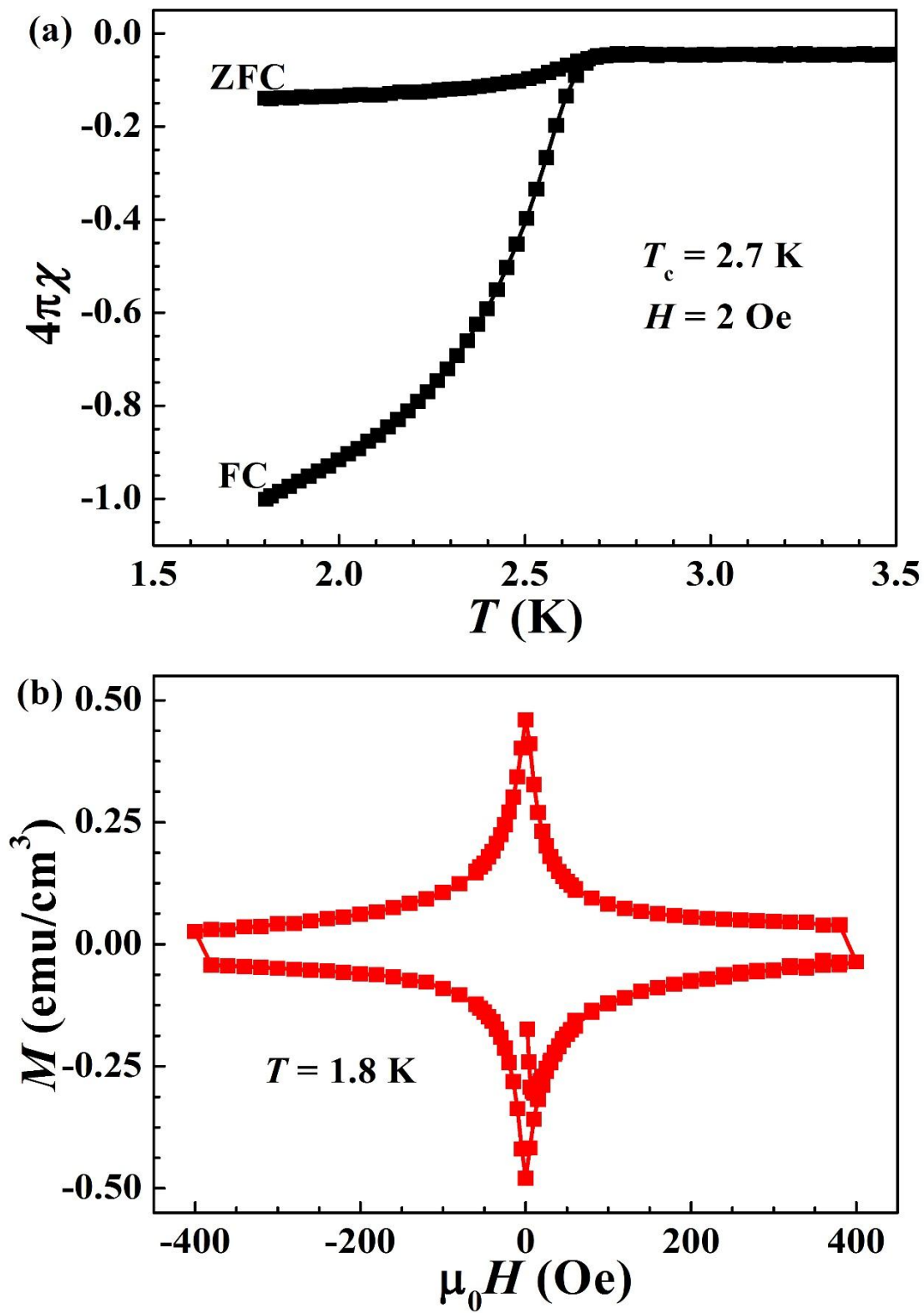

**Table 1:**

| Atom | X | Y | Z | Occ |
|------|------|------|-----------|-----------|
| La | 0.25 | 0.25 | 0.7305(7) | 1 |
| Pd1 | 0.25 | 0.25 | 0.1043(2) | 0.6038(6) |
| Bi1 | 0.25 | 0.25 | 0.3455(8) | 1 |
| Pd2 | 0.75 | 0.25 | 0.5 | 1 |
| Bi2 | 0.75 | 0.25 | 0 | 1 |